\documentclass[10pt,twocolumn,aps,pre,superscriptaddress]{revtex4-2}
    \usepackage[utf8]{inputenc}
    \usepackage{physics}
    \usepackage{amsmath}
    \usepackage{amsthm}
    \usepackage{amssymb}
    \usepackage{siunitx}
    \usepackage{hyperref}
    \usepackage{graphicx}
    \usepackage{tikz}

    \newcommand{\me}{\mathrm{e}}

    \makeatletter

    \theoremstyle{plain}
    \newtheorem{thm}{\protect\theoremname}
    \theoremstyle{plain}
    \newtheorem{prop}[thm]{\protect\propositionname}
    \makeatother
    
    \providecommand{\propositionname}{Proposition}
    \providecommand{\theoremname}{Theorem}

\begin{document}
    \title{Qubit Reset with a Shortcut-to-Isothermal Scheme}
    \author{Hong-Bo Huang}
        \affiliation{Graduate School of China Academy of Engineering Physics, No.10 Xibeiwang
        East Road, Haidian District, Beijing, 100193, China}
    \author{Geng Li}
        \affiliation{Graduate School of China Academy of Engineering Physics, No.10 Xibeiwang East Road, Haidian District, Beijing, 100193, China}
        \affiliation{School of Systems Science, Beijing Normal University, Beijing 100875,
        China}
    \author{Hui Dong}
        \email{hdong@gscaep.ac.cn}
        \affiliation{Graduate School of China Academy of Engineering Physics, No.10 Xibeiwang East Road, Haidian District, Beijing, 100193, China}
    \date{\today}
    \begin{abstract}
        Landauer's principle shows that the minimum energy cost to reset a classical bit in a bath with temperature $T$ is $k_{B}T\ln2$ in the infinite time. However, the task to reset the bit in finite time has posted a new challenge, especially for quantum bit (qubit) where both the operation time and controllability are limited. We design a shortcut-to-isothermal scheme to reset a qubit in finite time $\tau$ with limited controllability. The energy cost is minimized with the optimal control scheme with and without nonholonomic constraint. This optimal control scheme can provide a reference to realize qubit reset with minimum energy cost for the limited time.
    \end{abstract}
    \maketitle
    \section{Introduction\label{SecIntr}}
        Quantum information and quantum computation, a frontier interdisciplinary field of quantum mechanics and computer or information science, is developing rapidly in recent years. One of its goals is to realize the quantum computer to store, process and transmit information to complete the tasks that cannot be completed by traditional classical methods\cite{nielsen2002quantum}. Quantum bit (qubit) is the basic unity for quantum computer to store, process and transmit information\cite{nielsen2002quantum}. Different with classical computer, the availability qubits are typically limited due to the difficulties of the manufacture. Resetting qubit for reuse is therefore an inevitable step\cite{landauer1961irreversibility,bennett1973logical,zurek1989thermodynamic,bennett1997capacities,nielsen2002quantum,parrondo2015thermodynamics} for the qubit-demanding tasks.
        \par
        The process of bit resetting is to restore its state irreversibly to one particular state regardless of its initial state. Such an irreversibility process will leads to unavoidable energy cost. Landauer derived the famous \textit{Landauer's Principle} in 1961\cite{landauer1961irreversibility}, to reset $\SI{1}{\bit}$ in a heat bath with temperature $T$ requires a minimum energy cost of $k_{B}T\ln2$\cite{landauer1961irreversibility,berut2012experimental,jun2014high,peterson2016experimental,yan2018single,dago2021information}. Such bound is reached only in an infinite-time process. However, we are limited by the available operation time in the quantum computation process, which should be completed within the coherence time of the qubit\cite{nielsen2002quantum}. Such new scenario has posted a quest for the quantum generalization of the Landauer's principle for the finite-time reset processes\cite{nielsen2002quantum,dago2021information,ma2020experimental,ma2022minimal,diana2013finite,zulkowski2014optimal,zulkowski2015optimal,proesmans2020finite,proesmans2020optimal,van2022finite,feynman1985quantum,divincenzo1995quantum,divincenzo2000physical,zhen2021universal,rolandi2022finite,miller2023finite}.
        \par
        One of the possible protocol is to use the shortcut-to-isothermal scheme, where the system is driven with an auxiliary Hamiltonian $H_{\text{a}}$ in additional to the original Hamiltonian $H_{\text{o}}$ to evolve along the instantaneous equilibrium state of $H_{\text{o}}$ within a finite-time process\cite{li2017shortcuts,li2022geodesic}. Such scheme has been applied into many fields to reduce the time to drive the system from one state to another\cite{albay2019thermodynamic,albay2020realization,albay2020work}, to control biological evolution\cite{iram2021controlling,ilker2022shortcuts}, to construct finite-time engines\cite{quan2007quantum,pancotti2020speed,nakamura2020fast,plata2020building}, and to improve the accuracy of free-energy estimation\cite{li2021equilibrium}. Such finite-time process typically accompanies with additional energy consumption due to irreversibility produced in the finite-time thermodynamical process. In the current application into qubit reset, the controllability of the quantum devices also limits the available auxiliary Hamiltonian $H_{\text{a}}$ in the shortcut-to-isothermal scheme. Therefore, we consider the shortcut-to-isothermal reset process of qubit for the condition with and without the limits of the controllability, referred as bounded and unbounded control condition in the later discussion.
        \par
        The rest of the current paper is organized as follow. In Sec. \ref{SecUnbCtrl}, we introduce the concept of the qubit reset, design a shortcut-to-isothermal scheme on the qubit reset and find the optimal control protocol to minimize the extra work in the unbounded control condition. In Sec. \ref{SecBCtrl}, we obtain the optimal scheme for the bounded control condition, and show the cases where the experimental conditions are regarded as bounded or unbounded. And we also show the existence of the inaccessible region where the desired reset state is not achievable due to experimental limitations. The extra work is also calculated for the bounded control condition and compared with the bounded control condition. In Sec. \ref{SecConl}, we conclude the paper with additional discussions.
    \section{Minimum Energy Cost for Unbounded Control\label{SecUnbCtrl}}
        For the quantum computing devices, qubit is the fundamental element whose two quantum states are encoded as the logical states $0$ and $1$ respectively. The evolution is described by the Hamiltonian 
        \begin{align}
            H_{\text{o}}\bqty{\lambda\pqty{t}}=\frac{1}{2}\lambda\pqty{t}\sigma_{z}\text{,}
        \end{align}
        where the excited state $\ket{e}$ represents the logical state $1$ and the ground state $\ket{g}$ represents the logical state $0$. $\sigma_{z}=\ketbra{e}-\ketbra{g}$ is the Pauli operator. And the energy difference $\lambda\pqty{t}$ between the two states is tuned by an outside agent under a given protocol to realize the reset process.
        \par
\begin{figure}
            \includegraphics{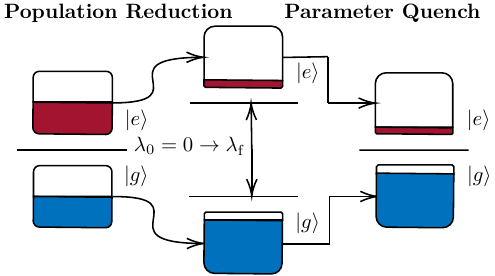}
            \caption{Two steps to reset a qubit. In the first step, the energy difference is raised from $\lambda_{0}=0$ to $\lambda_{\text{f}}$ within the heat bath at the temperature $T$ to reduce the population of excited state $\ket{e}$ with a small reset error $\epsilon$ ($\lambda_{\text{f}}\to\infty,\epsilon\to0$ for ideal reset). In the second step, we reset the energy difference and keep the population unchanged.\label{Fig2Stp}}
        \end{figure}

        The process to reset a qubit is to drive its state evolution to the ground state $\ket{g}$ despite its initial state. A straightforward scheme is to increase the energy difference to drive the major population to the ground state. Let $p_{e}\pqty{t},p_{g}\pqty{t}$ be the population of the excited state and the ground state respectively. The goal of the reset process typically lies in two aspects. The first one is to reduce the population $p_{e}$ on the excited state to a tolerable precision, \textit{i.e.}, $p_{e}\pqty{\tau}=\epsilon$ in the finite time $\tau$. The second one is to reset the control parameter to its original value, \textit{i.e.}, $\lambda\pqty{\tau}=\lambda\pqty{0}$, for subsequent operations. To achieve the goal above, we design a two-step scheme, illustrated in Fig. \ref{Fig2Stp}.
        \begin{enumerate}
            \item \textbf{Population Reduction with the shortcut-to-isothermal scheme.} Raise the energy difference from $\lambda_{0}$ to $\lambda_{\text{f}}$ to reduce the population of excited state $\ket{e}$. Such population reduction is done by the shortcut-to-isothermal scheme presented later. It worth mention that our shortcut scheme ensures that the bit is reset to logical $0$ with a small error
            \begin{align}
                \epsilon=\frac{\me^{-\beta\lambda_{\text{f}}}}{1+\me^{-\beta\lambda_{\text{f}}}}.
            \end{align}
            \item \textbf{Parameter Quench.} Reset the energy difference and keep the population unchanged. In this step, the energy difference of the system is reset to $\lambda_{0}$ and the population remains unchanged.
        \end{enumerate}
        By the end of the two steps, not only is the population reset to logical $0$, but also the system's parameter is reset to the initial value.
        \begin{figure}
            \includegraphics{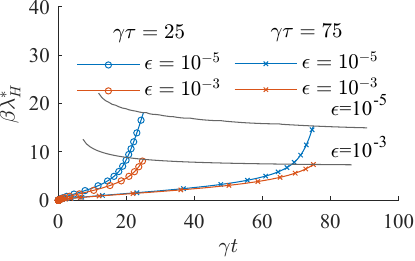}
            \caption{The optimal control scheme $\lambda_{H}^{*}$ (in units of $\beta^{-1}$) as the function of time $t$ (in units of $\gamma^{-1}$) with different reset times $\tau$ (in units of $\gamma^{-1}$). The numerical simulation is performed by shooting method and the boundary condition is $p_{e}\pqty{0}=1/2$ and $p_{e}\pqty{\tau}=\epsilon$. Different colors represent different reset errors $\epsilon=10^{-3}$ (red) and $10^{-5}$ (blue) and different markers represent difference reset times $\gamma\tau=25$ (circle) and $75$ (cross) respectively. The gray dashed lines show the final control amplitude $\lambda_{H}\pqty{\tau}$ given reset error $\epsilon$. It is clear to observe that for the optimal control scheme, the control parameter $\lambda_{H}^{*}\pqty{t}$ monotonically increases, and for fixed reset error larger control amplitude $\lambda_{H}\pqty{\tau}$ is needed for shorter reset time $\tau$.\label{FigSt}}
        \end{figure}
        \par
        In the normal control process, the evolution of the system typically has a lag, which prevents the qubit to reach the desired control precision. To overcome such lag, we introduce the shortcut-to-isothermal to escort the system evolution along the designed path. In the shortcut-to-isothermal scheme, an auxiliary Hamiltonian $H_{\text{a}}$ is added into the original Hamiltonian to escort the system evolution as the instantaneous equilibrium states $\rho_{\text{sc,eq}}^{\pqty{1}}\bqty{\lambda\pqty{t}}=\exp\pqty{-\beta H_{\text{o}}\pqty{t}}/\tr\pqty{\exp\pqty{-\beta H_{\text{o}}\pqty{t}}}$ of the original Hamiltonian $H_{\text{o}}$. Here $\beta=\pqty{k_BT}^{-1}$ is the inverse temperature with Boltzmann constant $k_{B}$. Normally, we ensure the vanish of the current auxiliary Hamiltonian $H_{\text{a}}\pqty{0_{-}}=H_{\text{a}}\pqty{\tau_{+}}=0$ to remove the additional control.
        \par
        A straightforward auxiliary Hamiltonian for the qubit is
        \begin{align}
            H_{\text{a}}\bqty{\lambda_{\text{a}}\pqty{t}}=\frac{1}{2}\lambda_{\text{a}}\pqty{t}\sigma_{z}\text{.}
        \end{align}
        The total Hamiltonian is $H_{\text{tot}}=1/2\lambda_{H}\pqty{t}\sigma_{z}$, where $\lambda_{H}\pqty{t}=\lambda\pqty{t}+\lambda_{\text{a}}\pqty{t}$ is an effective energy gap. The master equation of motion for the two-level system is\cite{breuer2002theory}
        \begin{align}
            &\dv{t}\rho\pqty{t}\nonumber \\
            =&-\frac{\gamma}{2}\pqty{\sigma_{-}\sigma_{+}\rho\pqty{t}-2\sigma_{+}\rho\pqty{t}\sigma_{-}+\rho\pqty{t}\sigma_{-}\sigma_{+}}n\pqty{\lambda_{H}}\nonumber \\
            &-\frac{\gamma}{2}\pqty{\sigma_{+}\sigma_{-}\rho\pqty{t}-2\sigma_{-}\rho\pqty{t}\sigma_{+}+\rho\pqty{t}\sigma_{+}\sigma_{-}}\pqty{n\pqty{\lambda_{H}}+1}\text{,}\label{EqMEq}
        \end{align}
        where $\sigma_{+}=\ketbra{e}{g},\sigma_{-}=\ketbra{g}{e}$. Here $n\pqty{\lambda_{H}}=\pqty{\exp\pqty{\beta\lambda_{H}}-1}^{-1}$ is the average phonon number for the thermal bath mode with the frequency $\lambda_{H}$.
        \begin{figure}
            \includegraphics{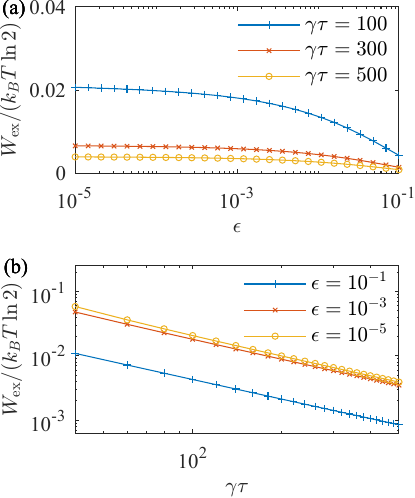}
            \caption{(a) The minimum extra work $W_{\text{ex}}$ (in units of $\beta^{-1}\ln2$) vs the reset error $\epsilon$ in linear-log plot. Different colors and markers represent difference reset times $\gamma\tau=100$ (blue plus sign), $300$ (red cross) and $500$ (yellow circle) respectively. We can see that the lower reset error we desire, the higher extra work we need to down. And when the reset error $\epsilon\to 0$, the extra work approaches to a constant. (b) The minimum extra work $W_{\text{ex}}$ (in units of $\beta^{-1}\ln2$) vs the reset time $\tau$ (in units of $\gamma^{-1}$) in log-log plot. A line with a slope of $-1$, shows the inverse relation between the minimum extra work $W_{\text{ex}}$ and the reset time $\tau$. Different colors and markers represent difference reset errors $\epsilon=10^{-1}$ (blue plus sign), $10^{-3}$ (red cross) and $10^{-5}$ (yellow circle) respectively.\label{FigWext}}
        \end{figure}
        \par
        With the instantaneous equilibrium state $\rho\pqty{t}=\rho_{\text{sc,eq}}^{\pqty{1}}\bqty{\lambda\pqty{t}}$, the evolution of the excited state population is given by the following equation
        \begin{align}
            \dv{p_e}{t}=\gamma\frac{\me^{-\beta\lambda_{H}}\pqty{1-p_{e}}-p_{e}}{1-\me^{-\beta\lambda_{H}}}\text{.}\label{EqPtoT}
        \end{align}
        \par
        To find the optimal control, we firstly calculate the energy cost for the two-step resetting process. The energy cost in the first step $W_{\text{sc}}^{\pqty{1}}=\int_{0}^{\tau}\tr\pqty{\rho_{\text{sc,eq}}^{\pqty{1}}\dot{H}}\dd{t}$\cite{quan2007quantum,alicki1979quantum} is obtained explicitly as
        \begin{align}
            W_{\text{sc}}^{\pqty{1}}=&\frac{1}{\beta}J+\lambda_{H}\pqty{\tau}\pqty{\epsilon-\frac{1}{2}}\text{,}
        \end{align}
        where $J=-\gamma\beta\int_{0}^{\tau}\pqty{\me^{-\beta\lambda_{H}}\pqty{1-p_{e}}-p_{e}}/\pqty{1-\me^{-\beta\lambda_{H}}}\lambda_{H}\mathrm{d}t$. In the second step, the state of the system remains unchanged $\rho_{\text{qa}}^{\pqty{2}}=\rho_{\text{sc,eq}}^{\pqty{1}}\bqty{\lambda_{\mathrm{f}}}$. The energy cost is obtained as\cite{ma2022minimal}
        \begin{align}
            W_{\text{qa}}^{\pqty{2}}=-\lambda_{H}\pqty{\tau}\pqty{\epsilon-\frac{1}{2}}\text{.}
        \end{align}
        And the total energy cost in our shortcut-to-isothermal scheme $W_{\text{sc}}=W_{\text{sc}}^{\pqty{1}}+W_{\text{qa}}^{\pqty{2}}$ is
        \begin{align}
            W_{\text{sc}}=\frac{1}{\beta}J\text{.}
        \end{align}
        The energy to reset the bit for quasi-static process $W_{\text{qs}}$ is the change of the free energy\cite{ma2022minimal}
        \begin{align}
            W_{\text{qs}}=\Delta{F}=\frac{1}{\beta}\pqty{\ln2-S\pqty{\epsilon}}\text{,}
        \end{align}
        where $S\pqty{\epsilon}=-\epsilon\ln\epsilon-\pqty{1-\epsilon}\ln\pqty{1-\epsilon}$ is the Shannon entropy of the final state. For ideal reset $\epsilon\to 0$, $S\pqty{\epsilon}\to 0$, the energy cost reached the Landauer's limit. Here we aim to find the extra cost of energy to reset the qubit due to finite-time operation. The extra energy $W_{\text{ex}}=W_{\text{sc}}-W_{\text{qs}}$ is obtained explicitly as
        \begin{align}
            W_{\text{ex}}=\frac{1}{\beta}\pqty{J-\pqty{\ln2-S\pqty{\epsilon}}}\text{.}\label{EqWex}
        \end{align}
        For the fixed reset error $\epsilon$, the task to minimize the extra work $W_{\text{ex}}$ is converted to the question to find the minimum of the objective function $J$ with constraints as follows,
        \begin{align*}
            &\text{equation of motion:}&&\dv{p_e}{t}=\gamma\frac{\me^{-\beta\lambda_{H}}\pqty{1-p_{e}}-p_{e}}{1-\me^{-\beta\lambda_{H}}}\\
            &\text{objective function:}&&J\bqty{p_{e}\pqty{t};\lambda_{H}\pqty{t}}\\
            &&&=-\gamma\beta\int_{0}^{\tau}\frac{\me^{-\beta\lambda_{H}}\pqty{1-p_{e}}-p_{e}}{1-\me^{-\beta\lambda_{H}}}\lambda_{H}\dd{t}\\
            &\text{boundary conditions:}&&\begin{cases}
                p_{e}\pqty{0}=\frac{1}{2}\\
                p_{e}\pqty{\tau}=\epsilon
            \end{cases}
        \end{align*}
        \par
        We introduce an effective Lagrange $L\pqty{p_{e};\lambda_{H}}=-\gamma\beta\pqty{\me^{-\beta\lambda_{H}}\pqty{1-p_{e}}-p_{e}}/\pqty{1-\me^{-\beta\lambda_{H}}}\lambda_{H}$. The cost function is rewritten as $J=\int_{0}^{\tau}L\dd{t}$. And the minimum is obtained by solving the Euler-Lagrange equation $\partial{L}/\partial{p_{e}}-\dd\pqty{\partial{L}/\partial{\dot{p}_{e}}}/\dd{t}=0$, which yields a 2nd-order ordinary differential equation for $p_{e}\pqty{t}$ as follows,
        \begin{align}
            \ddot{p}_{e}=\frac{\gamma^{2}\pqty{1-2p_{e}+2p_{e}^{2}}\dot{p}_{e}^{2}+2\gamma\dot{p}_{e}^{3}+2\dot{p}_{e}^{4}}{\gamma\pqty{1-2p_{e}}\pqty{2\gamma p_{e}\pqty{1-p_{e}}+\dot{p}_{e}}}.\label{EqP}
        \end{align}
        Here, we have used the equation of motion to replace the parameter $\lambda_{H}$. The solution of the above equation is denoted as $p_{e}^{*}\pqty{t}$. Substituting $p_{e}^{*}\pqty{t}$ into equation of motion, we can get the corresponding control scheme $\lambda_{H}^{*}\pqty{t}$.
        \par
        In Fig. \ref{FigSt}, we show the optimal control scheme $\lambda_{H}^{*}\pqty{t}$ with different reset time $\tau$. The numerical simulation is performed by solving Eq. (\ref{EqP}) with the shooting method\cite{li2022geodesic}. The boundary conditions are $p_{e}\pqty{0}=1/2$ and $p_{\tau}=\epsilon$. In the simulation, we have chosen different reset errors $\epsilon=10^{-1}$ and $10^{-5}$ under different reset times $\gamma\tau=25$ and $75$ respectively. The gray dashed lines show the final control amplitude $\lambda_{H}\pqty{\tau}$ given reset error $\epsilon$.
        \par
        With these curves, we observe two facts as follows,
        \begin{itemize}
            \item For the optimal control scheme, the control parameter $\lambda_{H}^{*}\pqty{t}$
            increases monotonically with time $t$.
            \item For fixed reset error, larger control amplitude $\lambda_{H}\pqty{\tau}$
            is needed for shorter reset time $\tau$.
        \end{itemize}
        With the optimal control scheme $\lambda_{H}^{*}$, we calculate the extra work $W_{\text{ex}}$. In Fig. \ref{FigWext}, we show the extra work $W_{\text{ex}}$ as functions of the reset error $\epsilon$ in Fig. \ref{FigWext} (a) and the control time $\tau$ in Fig. \ref{FigWext} (b). In Fig. \ref{FigWext} (a), different colors and markers represent difference reset times $\gamma\tau=$$100$ (blue plus sign), $300$ (red cross) and $500$ (yellow circle) respectively. For fixed reset time, the lower the reset error we desire, the larger the extra work is needed. For $\epsilon\to0$, the extra work $W_{\text{ex}}$ approaches to a constant. In Fig. \ref{FigWext}(b), different colors and markers represent different reset errors $\epsilon=$$10^{-1}$ (blue plus sign), $10^{-3}$ (red cross) and $10^{-5}$ (yellow circle) respectively. The curves show that the extra work is inversely proportional to reset time for fixed reset error.
        \par
        Noticing the second fact, we may find the control parameter $\lambda_{H}$ reaches the maximum value of the available control range, \textit{i.e.} $\lambda_{H}\in\bqty{0,\lambda_{m}}$ in some situations, which is discussed in the next section.
    \section{Minimum Energy Cost for Bounded Control\label{SecBCtrl}}
        In this section, we consider the energy minimization in bounded control condition $\lambda_{H}\in\bqty{0,\lambda_{m}}$, which is determined by the detailed condition in the experimental setup. Taking transmon superconducting qubit in quantum harmonic oscillator with Josephson junction as an example, the energy difference is controlled by changing the Josephson energy $E_{J}$, the capacitive energy $E_{C}$, and the inductive energy $E_{L}$\cite{koch2007charge,kjaergaard2020superconducting,devoret2013superconducting,oliver2013materials,gu2017microwave,bader2013transmon,huang2020superconducting,yin2015quantum}. For a single-junction transmon superconducting qubit, the energy difference is set by the size of the shunt capacitor and the critical current of the Josephson junction $I_{C}$, determined by design and fabrication parameters such as materials choice, junction area, and insulator thickness\cite{kjaergaard2020superconducting}. By replacing the single Josephson junction by a superconducting loop with two junctions in parallel---a dc-SQUID, the energy difference can be tuned because the effective critical current $I_{C}$ of the Josephson junction can be tuned via a magnetic field applied to dc-SQUID. Therefor, the limit of magnetic flux on dc-SQUID limits the energy difference\cite{wellstood1987low,kjaergaard2020superconducting}. And naturally, the technological limits of Josephson junction, capacitor and inductor also limit the energy difference\cite{devoret2013superconducting,gu2017microwave}. The typical experimental parameter is $\lambda_{m}\sim 2\pi\times \SI{10}{\giga\hertz}\ \text{GHz}$ or $\beta\lambda_{m}\sim5$ for superconducting qubit's typical working temperature $T=\SI{10}{\milli\kelvin}$\cite{gu2017microwave}. Such limitation results in a different control scheme as shown in the later discussion.
        \par
        To simplify the discussion, we firstly introduce a proposition for the bounded control condition with its proof presented in the Appendix \ref{AppxPT1}.
        \begin{prop}
            For the optimal reset scheme, if $\lambda_{H}$ touches the upper boundary \textup{$\lambda_{m}$} at the touch time $t^{*}$, then $\lambda_{H}$ will remain $\lambda_{m}$ for later time $t^{*}<t<\tau$ in the control protocol.\label{prop1}
        \end{prop}
        \par
        With this proposition, we divide the control time $\tau$ into three possible cases for given reset precision $\epsilon$ as inaccessible, untouched and touched\cite{plata2019optimal} as
        \begin{figure}
            \includegraphics{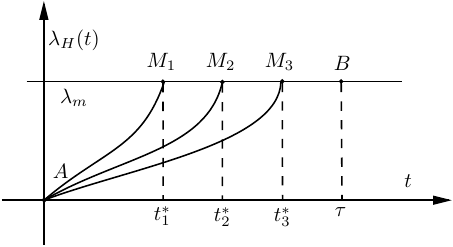}
            \caption{The optimization of the touch time $t^{*}$. All the curves $AM_{1},AM_{2}$ and $AM_{3}$ are the solution of Euler-Lagrange equation $\lambda_{H}^\pqty{\text{opt}}\pqty{t}=\lambda_{H}^{*}\pqty{t}$ and all the lines $M_{1}B,M_{2}B$ and $M_{3}B$ are on the boundary $\lambda_{H}^\pqty{\text{opt}}\pqty{t}=\lambda_{H}^{*}\pqty{t}$. Different points $M_{1},M_{2}$ and $M_{3}$ represent different touch time $t_{1}^{*},t_{2}^{*}$ and $t_{3}^{*}$ respectively. It is still to be considered that which $M\pqty{t^*}$ is the optimal. It is proved that the touch time $t^{*}$ is optimized by the continuity of $\lambda_{H}^\pqty{\text{opt}}$ and $p_{e}^\pqty{\text{opt}}$in Appendix \ref{AppxCCt}.\label{FigOptT}}
        \end{figure}
        \par
        \begin{enumerate}
            \item \textbf{Inaccessible.} There exists a critical reset $\tau_{c1}$, named as the first critical time. Once the reset time $\tau<\tau_{c1}$, one can not drive the system to the final state with the precision $\epsilon$ even with $\lambda_{H}\pqty{t}=\lambda_{m}$ for all time $t\in\bqty{0,\tau}$. The explicit form $\tau_{c1}$ is presented in the Appendix \ref{AppxCRT}.
            \item \textbf{Untouched.} Once the reset time $\tau>\tau_{c2}$, the final parameter $\lambda_{H}\pqty{\tau}$ is smaller than $\lambda_{m}$, \textit{i.e.}, $\lambda_{H}\pqty{\tau}<\lambda_{m}$. Here, we define $\tau_{c2}$ as the second critical reset time, which depends on the reset error $\epsilon$. The optimal reset process is the same as that in the unbounded control condition.
            \item \textbf{Touched.} For the control time $\tau_{c1}<\tau<\tau_{c2}$, there exists a touch time $t^{*}$ with $\lambda_{H}^{*}\pqty{t^*}=\lambda_{m}$. The optimal control scheme is obtained as follows,
            \begin{align}
                \lambda_{H}^\pqty{\text{opt}}\pqty{t}=\begin{cases}
                    0&t=0_{-}\text{,}\\
                    \lambda_{H}^{*}\pqty{t}&0<t<t^{*}\text{,}\\
                    \lambda_{m}&t^{*}<t<\tau\text{,}\\
                    0&t=\tau_{+}\text{.}
                \end{cases}\label{EqLHbd}
            \end{align}
            Here the touch time $t^{*}$ is also a variable to be optimized, which is illustrated in Fig. \ref{FigOptT}. All the curves $AM_{1},AM_{2}$ and $AM_{3}$ are the solution of Euler-Lagrange equation $\lambda_{H}^\pqty{\text{opt}}\pqty{t}=\lambda_{H}^{*}\pqty{t}$ and all the lines $M_{1}B,M_{2}B$ and $M_{3}B$ are on the boundary $\lambda_{H}^\pqty{\text{opt}}\pqty{t}=\lambda_{H}^{*}\pqty{t}$. Different points $M_{1},M_{2}$ and $M_{3}$ represent different touch times $t_{1}^{*},t_{2}^{*}$ and $t_{3}^{*}$ respectively. It is still to be considered that which $M\pqty{t^*}$ is the optimal. It is proved that the touch time $t^{*}$ is optimized by the continuity of $\lambda_{H}^\pqty{\text{opt}}$ and $p_{e}^\pqty{\text{opt}}$in Appendix \ref{AppxCCt}.
        \end{enumerate}
        \begin{figure}
            \includegraphics{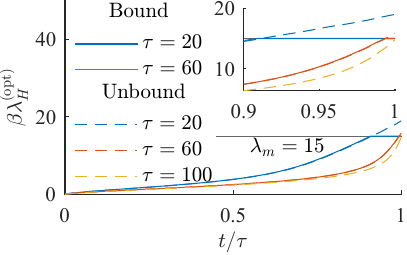} 
            \caption{The optimal control scheme with (solid line) and without (dashed line) bound as the function of normalized time $t/\tau$. The gray line is the bound $\beta\lambda_{m}=15$. Resit error is set to $\epsilon=10^{-5}$. In this situation, $\gamma\tau_{c1}=10.9$ and $\gamma\tau_{c2}=94.1$. Different colors represent difference reset times $\gamma\tau=20$ (blue), $60$ (red) and $100$ (yellow) respectively. The subplot is indicates the area magnified in interval $t/\tau\in\bqty{0.9,1}$. At the bound control condition, the optimal scheme stay on the boundary once it touches the boundary at $t^{*}$. For $\tau=100\geq\tau_{c_{2}}$, the bound control condition is the same as the unbound control condition so only the solid line is drawn in the figure.\label{FigBd}}
        \end{figure}
        \par
        Fig. \ref{FigBd} shows the optimal control schemes with and without bound for different control time $\tau$. The parameters are set as bound $\beta\lambda_{m}=15$, resit error $\epsilon=10^{-5}$. The solid line and dashed line represents the optimal control scheme with and without bound respectively. For the bounded control, the optimal scheme (the solid line) $\lambda_{H}^{*}$ touches the bound (the gray line) at the time $t=t^{*}$.
        \par
\begin{figure}
            \includegraphics{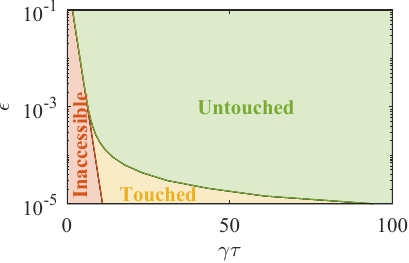}
            \caption{Cases diagram of the system in the $\pqty{\tau,\epsilon}$ linear-log plane for the bounded control problem. The red and the green lines respectively represent the boundary lines $\tau_{c1}$ and $\tau_{c2}$ of the three cases. The bound is set as $\beta\lambda_{m}=15$. The reset tasks which is represented by the point in red region can not be accomplished (inaccessible). The green region means that those reset tasks can be accomplished and do not touch the boundary (untouched). The yellow region represents those reset tasks which can be accomplished but touch the boundary (touched).\label{FigCas}}
        \end{figure}

        Fig. \ref{FigCas} shows the cases diagram of the system on $\pqty{\tau,\epsilon}$ linear-log plane. The green and red lines respectively represent the boundary lines $\tau_{c1}$ and $\tau_{c2}$ of the three cases. The parameter is set as $\beta\lambda_{m}=15$. The calculation of $\tau_{c1}$ and $\tau_{c2}$ is presented in Appendix \ref{AppxCRT}.
        \begin{figure}
            \includegraphics{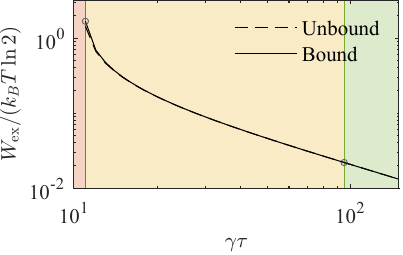}
            \caption{Minimum extra work done in the reset process as the function of reset time $\tau$ in log- log plot with (solid line) and without (dashed line) bound. The parameters are set as $\beta\lambda_{m}=15,\epsilon=10^{-5}$. The red and the green lines represent $\tau_{c1}$ and $\tau_{c2}$ respectively. The red, yellow and green regions represent the inaccessible, touched and untouched cases respectively. In the untouched region, the solid and the dashed lines are coincident because this case is untouched by the bound. In the touched region, the bound minimum extra work is larger than the unbound minimum extra work. And their difference increases with the decrease of the reset time. When $\tau\to\tau_{c1}$, the solid line tends to the maximal $W_{\text{ex,b,max}}$.\label{FigWexBd}}
        \end{figure}
        \par
        The extra work is calculated via the equation Eq.(\ref{EqWex}). For the touched case, the objective function is calculated in the two time intervals $\bqty{0,t^{*}}$ and $\bqty{t^{*},\tau}$ as
        \begin{align}
            J=J^{\pqty{1}}+J^{\pqty{2}}\text{,}
        \end{align}
        In the time interval $\bqty{0,t^{*}}$, the optimal control scheme is $\lambda_{H}^{\pqty{\text{opt}}}\pqty{t}\lambda_{H}^{*}\pqty{t},p_{e}^{\pqty{\text{opt}}}\pqty{t}=p_{e}^{*}\pqty{t}$, and the function is explicitly obtained as
        \begin{align}
            J^{\pqty{1}}=-\beta\gamma\int_{0}^{t^{*}}\frac{\me^{-\beta\lambda_{H}^{*}}\pqty{1-p_{e}^{*}}-p_{e}^{*}}{1-\me^{-\beta\lambda_{H}^{*}}}\lambda_{H}^{*}\dd{t}\text{.}
        \end{align}
        In the time interval $\bqty{t^{*},\tau}$, the optimal control scheme is $\lambda_{H}^{\pqty{\text{opt}}}\pqty{t}=\lambda_{m}$. Substituting it into the motion equation, we get
        \begin{align}
            p_{e}^{\pqty{\text{opt}}}\pqty{t}=&\frac{n\pqty{\lambda_{m}}}{2n\pqty{\lambda_{m}}+1}\nonumber \\
            & +\pqty{\epsilon-\frac{n\pqty{\lambda_{m}}}{2n\pqty{\lambda_{m}}+1}}\me^{-\gamma\pqty{2n\pqty{\lambda_{m}}+1}\pqty{t-\tau}}\text{,}
        \end{align}
        which is calculated in Appendix \ref{AppxCCt}. And $J^{\pqty{2}}$ is computed analytically
        \begin{align}
            J^{\pqty{2}}=&-\beta\gamma\int_{t^{*}}^{\tau}\frac{\me^{-\beta\lambda_{m}}\pqty{1-p_{e}^{\pqty{\text{opt}}}}-p_{e}^{\pqty{\text{opt}}}}{1-\me^{-\beta\lambda_{m}}}\lambda_{m}\mathrm{d}t\text{.}
        \end{align}
        For the untouched case, the extra work is the same as that in the unbound control condition.
        \par
        Fig. \ref{FigWexBd} shows the extra work with and without constraints. The meanings of the three colors are the same as in the cases diagram. In the yellow region, $W_{\text{ex,ub}}$ is larger than $W_{\text{ex,b}}$ and do not follow the inverse relation. And their difference increases with the decrease of the reset time.
        \par
        And when $\tau\to\tau_{c1}$, $J^{\pqty{1}}\to 0$ and the extra work tends to its maximal value $J_{\text{max}}=J_{\text{max}}^{\pqty{2}}$
        \begin{align}
            W_{\text{ex,b,max}}=\frac{1}{\beta}\pqty{J_{\text{max}}^{\pqty{1}}-\pqty{\ln2-S\pqty{\epsilon}}}\text{.}
        \end{align}
        where $J_{\text{max}}^{\pqty{2}}$ is obtained with $\tau\to\tau_{c1},t^{*}\to 0$.    
    \section{Conclusion\label{SecConl}}
        In this paper, we design a finite-time reset scheme based on the shortcut-to-isothermal approach, and find the optimal control scheme for the minimum extra energy cost with and without nonholonomic constraint $\lambda_{H}\in\bqty{0,\lambda_{m}}$. The scheme is a two-step scheme including population reduction and parameter quench. We find out the optional reset scheme $\lambda_{H}^{*}\pqty{t}$ and the extra energy cost $W_{\text{ex}}$ as the function of the reset error $\epsilon$ and the reset time $\tau$. The extra energy cost $W_{\text{ex}}$ follows the inverse proportional relationship as $W_{\text{ex}}\propto 1/\tau$.
        \par
        The bound on the controllability modify the problem of minimizing the extra work. In this condition, the system has three possible cases: inaccessible, untouched and touched. We show the existence of the first critical reset time $\tau_{c1}$ and the second critical reset time $\tau_{c2}$. When $\tau<\tau_{c1}$, the reset task is inaccessible. When $\tau_{c2}<\tau$, the reset task can be accomplished without any different with the unbound control condition. When $\tau_{c1}<\tau<\tau_{c2}$, the reset task can be accomplished but the optional reset scheme $\lambda_{H}^{\pqty{\text{opt}}}$ is not the same as $\lambda_{H}^{*}\pqty{t}$. And there exists a touch time $t^{*}$ to divide the optional reset scheme $\lambda_{H}^{\pqty{\text{opt}}}$ into two stage. First stage is $\lambda_{H}^{\pqty{\text{opt}}}\pqty{t}=\lambda_{H}^{*}\pqty{t}$ while $t<t^{*}$. Second stage is the boundary value $\lambda_{H}^{\pqty{\text{opt}}}\pqty{t}=\lambda_{m}$, while $t>t^{*}$.
    \begin{acknowledgments}
        This work is supported by the National Natural Science Foundation of China (NSFC) (Grants No. 12088101, No.U2230203).
    \end{acknowledgments}


    \appendix
    \section{the Prove of Proposition \ref{prop1}\label{AppxPT1}}
        The motion equation is rewritten as the equation of $\lambda_{H}$
        \begin{align}
            \lambda_{H}=-\beta\ln\frac{\dot{p}_{e}+\gamma p_{e}}{\dot{p_{e}}+\gamma\pqty{1-p_{e}}}\text{.}\label{AppxEqLtoP}
        \end{align}
        By differentiating Eq. (\ref{AppxEqLtoP}), we get
        \begin{align}
            \dv{\lambda_H}{t}=\frac{2\pqty{1-p_{e}\pqty{1+\me^{-\beta\lambda_{H}}}}\pqty{p_{e}-\me^{-\beta\lambda_{H}}\pqty{1-p_{e}}}}{\beta\pqty{1-2p_{e}}\pqty{p_{e}\pqty{1-\me^{-\beta\lambda_{H}}}+\me^{-\beta\lambda_{H}}}\pqty{1-\me^{-\beta\lambda_{H}}}}\text{.}\label{AppxEqDLtpP}
        \end{align}
        Noticing $0<\me^{-\beta\lambda_{H}}<1,0<p_{e}<1/2$, we observe that the following factors in the right side of Eq. (\ref{AppxEqDLtpP}) is lager than zero
        \begin{itemize}
            \item $\pqty{1-p_{e}\pqty{1+\me^{-\beta\lambda_{H}}}}>0$,
            \item $\pqty{1-2p_{e}}>0$,
            \item $\pqty{p_{e}\pqty{1-\me^{-\beta\lambda_{H}}}+\me^{-\beta\lambda_{H}}}>0$,
            \item $\pqty{1-\me^{-\beta\lambda_{H}}}>0$.
        \end{itemize}
        And with the factor $\pqty{p_{e}-\me^{-\beta\lambda_{H}}\pqty{1-p_{e}}}>0$ noticing $\lambda_{H}>\lambda$, we prove that $\lambda_{H}^{*}\pqty{t}$ is a monotonically increasing function with $\dd{\lambda_{H}}/\dd{t}>0.$
        \par
        If $\lambda_{H}^{*}$ touches the upper boundary $\lambda_{m}$ at the time $t^{*}$, it stay larger than $\lambda_{m}$ in the interval $\bqty{t^{*},\tau}$. Thus for the optional reset scheme, if the optimal control $\lambda_{H}^{\pqty{\text{opt}}}$ touches the upper boundary $\lambda_{m}$ at touch time $t^{*}$, $\lambda_{H}$stay on it for all times $t>t^{*}$.
    \section{the Two Critical Reset Time\label{AppxCRT}}
        The extreme case is $\lambda_{H}\pqty{t}=\lambda_{m}$ for the whole control process $t\in\bqty{0,\tau_{c1}}$ with the touch time $t^{*}=0$. With this control scheme, the motion equation becomes
        \begin{align}
            \dv{p_e}{t}=\gamma\pqty{-\pqty{2n\pqty{\lambda_{m}}+1}p_{e}+n\pqty{\lambda_{m}}},\label{AppxEqPtoNL}
        \end{align}
        whose solution is
        \begin{align}
            p_{e}\pqty{t}=\frac{n\pqty{\lambda_{m}}+\frac{1}{2}\me^{-\gamma\pqty{2n\pqty{\lambda_{m}}+1}t}}{2n\pqty{\lambda_{m}}+1}\overset{\text{def}}{=}\phi\pqty{t}\text{.}\label{AppxEqPhi}
        \end{align}
        The first critical reset time is obtained by setting $p_{e}\pqty{\tau_{c1}}=\epsilon$ as
        \begin{align}
            \tau_{c1}=\frac{1}{\gamma\pqty{2n\pqty{\lambda_{m}}+1}}\ln\frac{1}{2\pqty{\epsilon\pqty{2n\pqty{\lambda_{m}}+1}-n\pqty{\lambda_{m}}}}\text{.}
        \end{align}
        For any $\tau<\tau_{c1}$, the reset task with error $\epsilon$ can not be accomplished in the reset time $\tau$ .
        \par
        According to the Proposition \ref{prop1} in the main content, the condition for $\lambda_{H}^{*}$ not to touch the boundary is $\lambda_{H}^{*}\pqty{\tau}<\lambda_{m}$. The critical condition is that $\lambda_{H}^{*}$ touches $\lambda_{m}$ at the end of the control process $t=\tau$. And the second critical reset time $\tau_{c2}$ is given by $\lambda_{H}^{*}\pqty{\tau_{c2}}=\lambda_{m}$.
    \section{Using Continuity Condition to Calculate Touch Time $t^{*}$\label{AppxCCt}}
        In this section, we derive the condition for the optimal control in the touched case.
        \par
        We show that the $p_{e}^{\pqty{\text{opt}}}\pqty{t}$ and $\lambda_{H}^{\pqty{\text{opt}}}\pqty{t}$ is continuous at the optimal touch time $t^{*}$. The continuity of $p_{e}^{\pqty{\text{opt}}}\pqty{t}$ is natural because of the physical reason that the population is continuous. The motion equation Eq. (\ref{AppxEqLtoP}) ensures the continuity of the control scheme $\lambda_{H}^{\pqty{\text{opt}}}$ once $\dot{p}_{e}^{\pqty{\text{opt}}}$ is continuous in the whole process. The continuity of $\dot{p}_{e}^{\pqty{\text{opt}}}$ is proved as the result of the so-called one-sided variational problem as following.
        \par
        We write the objective function into two part with respect to the touch time $t^{*}$
        \begin{align}
            J\bqty{p_{e}\pqty{t}}=\int_{0}^{t_{-}^{*}}L\pqty{p_{e},\dot{p}_{e}}\mathrm{d}t+\int_{t_{+}^{*}}^{\tau}L\pqty{\phi,\dot{\phi}}\mathrm{d}t\label{AppxEqObJF}
        \end{align}
        In the second part, $p_{e}\pqty{t}$ is replaced with the defined function in Eq. (\ref{AppxEqPhi}) as $p_{e}\pqty{t}=\phi\pqty{t}$ to avoid the misunderstanding. The variation of the functional Eq. (\ref{AppxEqObJF}) is obtained as
        \begin{align}
            \delta J= & \eval{L\pqty{p_{e},\dot{p}_{e}}}_{t_{-}^{*}}\delta t^{*}+\int_{0}^{t_{-}^{*}}\pqty{\frac{\partial L}{\partial p_{e}}\delta p_{e}+\frac{\partial L}{\partial\dot{p}_{e}}\delta\dot{p}_{e}}\mathrm{d}t\nonumber \\
            & -\eval{L\pqty{\phi,\dot{\phi}}}_{t_{+}^{*}}\delta t^{*}\text{.}
        \end{align}
        With the integration by parts, we get
        \begin{align}
            \delta J= & \pqty{\eval{L\pqty{p_{e},\dot{p}_{e}}}_{t_{-}^{*}}-\eval{L\pqty{\phi,\dot{\phi}}}_{t_{+}^{*}}}\delta t^{*}\nonumber \\
            & +\eval{\frac{\partial L}{\partial\dot{p}_{e}}\delta p_{e}}_{t_{-}^{*}}+\int_{0}^{t_{-}^{*}}\pqty{\frac{\partial L}{\partial p_{e}}-\frac{\mathrm{d}}{\mathrm{d}t}\frac{\partial L}{\partial\dot{p}_{e}}}\delta p_{e}\mathrm{d}t\text{.}
        \end{align}
        The continuous condition $p_{e}\pqty{t_{-}^{*}}=\phi(t_{-}^{*})$ results in the $\delta p_{e}\pqty{t_{-}^{*}}=\eval{\dot{\phi}}_{t_{-}^{*}}\delta t^{*}$. Noticing $\delta p_{e}\pqty{t_{-}^{*}}=\eval{\delta p_{e}}_{t_{-}^{*}}+\eval{\dot{p}_{e}}_{t_{-}^{*}}\delta t^{*}$, we get $\eval{\delta p_{e}}_{t_{-}^{*}}=\eval{\pqty{\dot{\phi}-\dot{p}_{e}}}_{t_{-}^{*}}\delta t^{*}$. And the variation is simplified as
        \begin{align}
            \delta J= & \int_{0}^{t_{-}^{*}}\pqty{\frac{\partial L}{\partial p_{e}}-\frac{\mathrm{d}}{\mathrm{d}t}\frac{\partial L}{\partial\dot{p}_{e}}}\delta p_{e}\mathrm{d}t\label{AppxEqDObjF1}\\
            + & \pqty{\eval{\pqty{L\pqty{p_{e},\dot{p}_{e}}+\pqty{\dot{\phi}-\dot{p}_{e}}\frac{\partial L}{\partial\dot{p}_{e}}}}_{t_{-}^{*}}-\eval{L\pqty{\phi,\dot{\phi}}}_{t_{+}^{*}}}\delta t^{*}\text{.}\label{AppxEqDObjF2}
        \end{align}
        In the time interval $\bqty{0,t_{-}^{*}}$, the optimal control scheme ensures $p_{e}\pqty{t}=p_{e}^{*}\pqty{t}$ with the the Euler-Lagrange equation from Eq. (\ref{AppxEqDObjF1})
        \begin{align}
            \pdv{L}{p_e}-\dv{t}\pdv{L}{\dot{p}_{e}}=0\text{.}\label{AppxEqELEq}
        \end{align}
        At the time point $t^{*}$, we get the so-call transversality condition from Eq. (\ref{AppxEqDObjF2}) to connect the two part and point out the optional $t^{*}$
        \begin{align}
            -&L\pqty{\phi\pqty{t^{*}},\dot{\phi}\pqty{t^{*}}}+L\pqty{p_{e}^{*}\pqty{t^{*}},\dot{p}_{e}^{*}\pqty{t^{*}}}\nonumber\\
            &+\pqty{\dot{\phi}\pqty{t^{*}}-\dot{p}_{e}^{*}\pqty{t^{*}}}\frac{\partial L}{\partial\dot{p}_{e}}\pqty{\dot{p}_{e}^{*}\pqty{t^{*}},\dot{p}_{e}^{*}\pqty{t^{*}}}=0\text{.}
        \end{align}
        With the mean value theorem on the first two terms, we get
        \begin{align}
            \pqty{\dot{\phi}\pqty{t^{*}}-\dot{p}_{e}^{*}\pqty{t^{*}}}\pqty{\frac{\partial L}{\partial\dot{p}_{e}}\pqty{p_{e}^{*}\pqty{t^{*}},k}-\frac{\partial L}{\partial\dot{p}_{e}}\pqty{p_{e}^{*}\pqty{t^{*}},\dot{p}_{e}^{*}\pqty{t^{*}}}}\nonumber\\
            =0\text{,}
        \end{align}
        where $k$ is a value satisfying the condition $\min\pqty{\dot{p}_{e}^{*}\pqty{t^{*}},\dot{\phi}\pqty{t^{*}}}<k<\max\pqty{\dot{p}_{e}^{*}\pqty{t^{*}},\dot{\phi}\pqty{t^{*}}}$. Using the mean value theorem again, we get
        \begin{align}
            \pqty{\dot{\phi}\pqty{t^{*}}-\dot{p}_{e}^{*}\pqty{t^{*}}}\pqty{k-\dot{p}_{e}^{*}\pqty{t^{*}}}\frac{\partial^{2}L}{\partial\dot{p}_{e}^{2}}\pqty{p_{e}^{*}\pqty{t^{*}},l}=0\text{,}
        \end{align}
        where $l$ is a value satisfying the condition $\min\pqty{k,\dot{p}_{e}^{*}\pqty{t^{*}}}<k<\max\pqty{k,\dot{p}_{e}^{*}\pqty{t^{*}}}$.It is clear that
        \begin{align}
            \dot{\phi}\pqty{t^{*}}=\dot{p}_{e}^{*}\pqty{t^{*}}.
        \end{align}
        Therefore, $\lambda_{H}^{\pqty{\text{opt}}}$ is also continuous $\lambda_{H}^{*}\pqty{t^{*}}=\lambda_{m}$ because of the motion equation Eq. (\ref{AppxEqLtoP}).
        \par
        Explicitly, we get the optimal control as
        \begin{equation}
            p_{e}^{\pqty{\text{opt}}}\pqty{t}=\begin{cases}
                p_{e}^{*}\pqty{t}&0<t<t^{*}\text{,}\\
                \phi(t)&t^{*}<t<\tau\text{.}
            \end{cases}
        \end{equation}
        And the optional $t^{*}$ is obtained with the above condition $\lambda_{H}^{*}\pqty{t^{*}}=\lambda_{m}$.
    \bibliographystyle{apsrev4-2}
    \bibliography{Qubit_reset}
\end{document}